\documentclass[12pt]{article}
\usepackage{epsfig}
\usepackage{a4wide}
 
\newcommand{\kv}{{\bf k}}
\newcommand{\Qv}{{\bf Q}}
\def\Journal#1#2#3#4{{#1} {\bf #2}, #3 (#4)}

\def\PLB{{\em Phys. Lett.}  B}

\def\ZPC{{\em Z. Phys.} C}

\begin{document}

\rightline{TSL/ISV-2001-0249}
\rightline{June 2001}
\bigskip\bigskip
\begin{center}
{\Large \bf  Hard colour singlet exchange at the 
Tevatron\\[7pt] from the BFKL equation\protect\footnote{Talk 
presented by RE at the 9th International Workshop on Deep 
Inelastic Scattering, Bologna, April 2001, to appear in 
the proceedings. E-mail: enberg@tsl.uu.se.}} \\
\medskip
\bigskip\bigskip\bigskip

{{\bf R.~Enberg$^a$}, {\bf L.~Motyka$^{a,b}$}, {\bf G.~Ingelman$^{a,c}$}} \\
\medskip 
{$^a$ High Energy Physics, Uppsala University, Uppsala, Sweden}\\ 
{$^b$ Institute of Physics, Jagellonian University, Cracow, Poland}\\ 
{$^c$ DESY, Hamburg, Germany}\\ 
\bigskip
\end{center}

\vspace{.5cm}

\begin{abstract}
We study rapidity gaps between jets in $p\bar{p}$ collisions at the Tevatron
by a novel solution of the nonforward BFKL equation including nonleading effects
through the consistency constraint and running coupling. Results differ from
earlier calculations based on the asymptotic Mueller-Tang formula, but agree
with D0 data when including full event simulation using PYTHIA to model
the gap survival probability.
\end{abstract}
\bigskip
\vspace{.5cm}


\noindent
Mueller and Tang \cite{MT} (MT) proposed that events with two forward-backward
high-$E_T$ jets with a large rapidity gap (of size $y$) between them could 
provide a test
of the nonforward BFKL equation, in the form of elastic parton-parton
scattering by hard ($|t|\! \gg\! 0$) colour singlet exchange. However, the 
predictions
obtained for the $t$ and $y$ dependence are not confirmed by D\O\ \cite{D0} and
CDF \cite{CDF} data. To crudely reproduce the $t$-shape of the D\O\ data the
running of $\alpha_s(-t)$ had to be disregarded \cite{CFL}  and the $y$
dependence measured by CDF remains unexplained.  The MT approximation is valid
only for asymptotically large energies and neglects some contributions which
may be relevant for the present experimental conditions. Here, we show that a
more complete solution of the BFKL equation gives a good description of the
data.

The cross section $d\hat\sigma /dt$ for elastic parton-parton scattering via 
BFKL pomeron exchange is dominated by the imaginary part of the amplitude 
\begin{equation}
{\mathrm{Im}}\, A(y,t) = \int \frac{d^2\kv}{\pi} 
\frac{\Phi_0(\kv,\Qv)\Phi(y,\kv,\Qv)}
{[(\kv+\Qv/2)^2+s_0][(\kv-\Qv/2)^2+s_0]}, 
\end{equation}
\noindent where $\Qv/2\pm\kv$ denote the transverse momenta of the
exchanged gluons and $t=-Q^2$. The scale $s_0$ is a physical infrared cut-off
that reflects the confining properties of the QCD vacuum which suppresses the
propagation of gluons with small virtualities. We set $s_0$ to 1 GeV$^2$, but
note that the particular value affects mostly the normalisation and not the $y$
and $t$ dependence.

The impact factor $\Phi_0(\kv,\Qv)$ is  an
effective vertex between the pomeron, the proton and the 
final state containing the 
high $E_T$ parton(s). At large $|t|$ and $y$, the pomeron predominantly 
couples to individual pointlike partons \cite{Bartels}, corresponding to
\begin{equation}
\Phi_0(\kv,\Qv)\sim~\sqrt{\alpha_s((\Qv/2+\kv)^2+s_0)\, \alpha_s
((\Qv/2-\kv)^2+s_0)}.
\end{equation}
We assume that this is a good approximation also in the
nonasymptotic region. 
$\Phi(y,\kv,\Qv)$ is the gluon distribution including full $y$ evolution which
is found by solving the nonforward BFKL equation \cite{BFKL}
with  $\Phi_0(\kv,\Qv)$ as the input.
The nonleading corrections to the BFKL kernel are taken into account in an
approximate way by including the consistency constraint \cite{CC} and using a
running coupling where the scale $\mu^2 = k^2 +Q^2/4$ is evaluated locally in
the gluon ladder. 

\begin{figure}[t]
     \centerline{
 \epsfig{figure=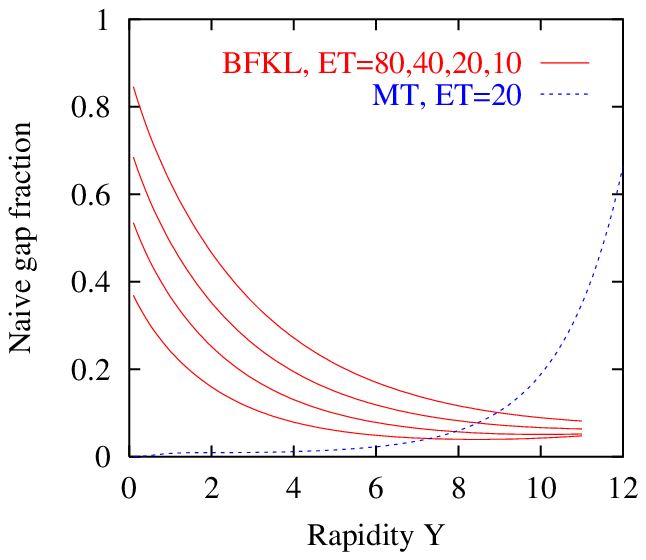, width=0.48\columnwidth
}
\epsfig{figure=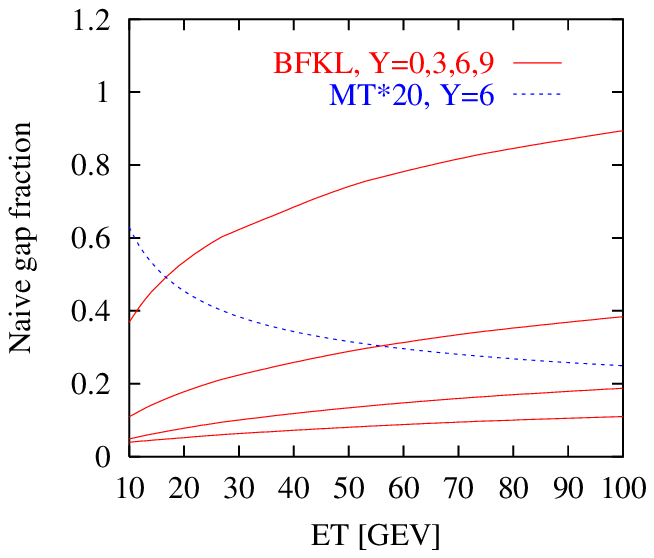, width=0.48\columnwidth
}
}
\caption{Parton level gap fractions (relative to LO QCD one-gluon exchange) in
quark-quark scattering from our BFKL solution and the asymptotic MT 
approximation
with the running coupling $\alpha_s(-t)$ in the prefactor, versus gap
size $y$ (for the given  $E_T$-values) and versus $E_T$ (for the given
$y$-values). In the right-hand plot, the MT curve is multiplied by a 
factor of 20.}\label{fig:parton}
\end{figure}

The BFKL equation is solved numerically by expanding it in a truncated
Fourier series and discretising the system by the Tchebyshev interpolation
method \cite{JKLM}. The resulting theoretical gap fraction is shown in
fig.~\ref{fig:parton} together with the result of the MT formula which has a
completely different behaviour. This is because MT corresponds to the
asymptotically leading part of the scattering amplitude, which is not
sensitive to the low scale details of the hadron structure and the QCD vacuum. 
However, the amplitude also contains a nonasymptotic part which depends on the 
physical cut-off scale and which dominates at small $y$. This part contains a
$\log({Q^2}/{s_0})$, visible at the Born level, which gets suppressed when $y$
increases. This is due to gluon reggeisation \cite{Bartels}, i.e., the negative
virtual correction terms in the BFKL kernel. 
The numerical value of $\log (Q^2/s_0)$ is large and the BFKL evolution is
not rapid enough to reach the high $y$ asymptotic limit in the currently 
accessible
domain of $y$. We stress, that this
$\log({Q^2}/{s_0})$ in not an infrared artifact, since it would also appear if
we considered the analogous process of heavy onia scattering, where $s_0$ 
($\Lambda_{QCD} \ll s_0\ll Q^2$) would be a hard scale corresponding to the
heavy onium size.

In order to compare with data, the parton level cross section discussed above
must be convoluted with QCD evolved parton distributions of the incoming
hadrons and the transition to final state hadrons be made. The latter is
particularly important when considering rapidity gaps, which are defined in
terms of a region without particles. One must then consider all additional
activity in an event,  such as higher order parton emissions, multiple parton
scattering and hadronisation. A simple way of doing this is to multiply the
cross section with an estimated gap survival probability. However, this
probability cannot be independent of the subprocess, with its associated colour
string topology, or of the kinematics in the event. For example, gluon-gluon
scattering and a larger hard scale resulting in more parton radiation is
expected to give a smaller gap survival probability.

\begin{figure}[t]
     \centerline{
         \epsfig{figure=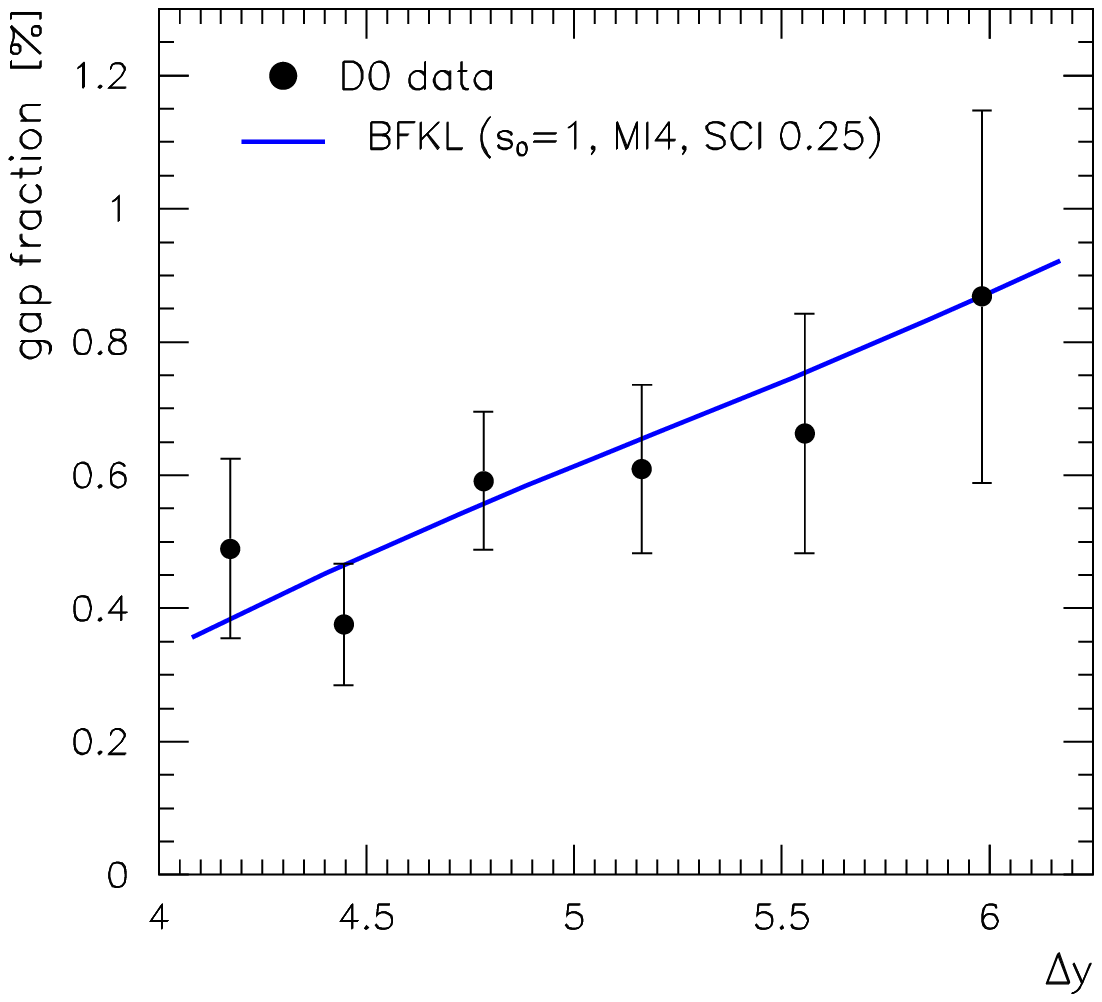, width=0.45\columnwidth, 
                 bburx=327,bbury=304,bbllx=5,bblly=13} 
         \epsfig{figure=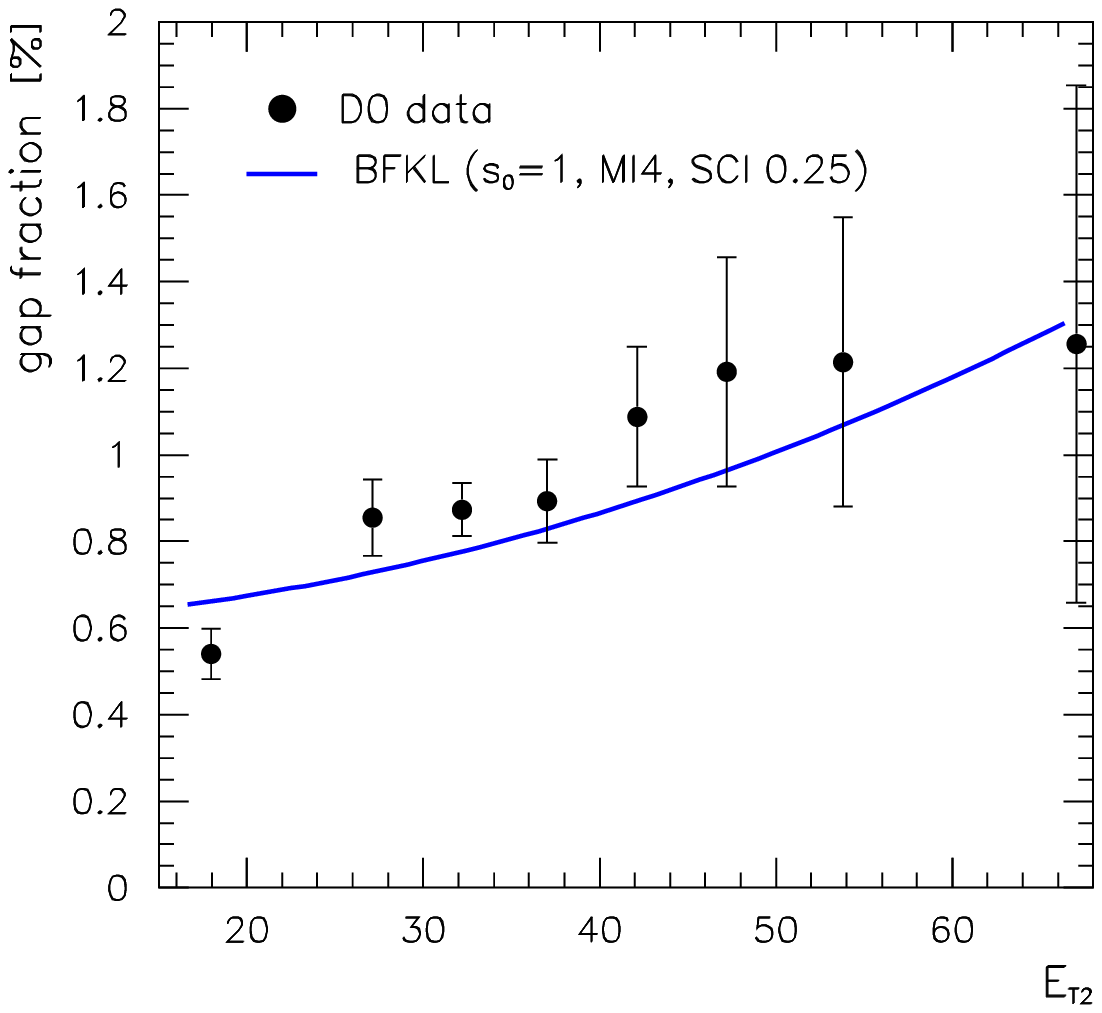, width=0.45\columnwidth,
                 bburx=327,bbury=304,bbllx=5,bblly=13} 
           }           
\vspace{-3mm}
\caption{Fraction of jet-gap-jet events over all jet events versus jet 
separation
$\Delta \eta$ for $15 < E_{T2} < 25$ GeV and $E_T$ of the lowest energy jet;
D\O\ data \protect\cite{D0} compared to our BFKL-based model implemented 
in {\sc pythia} relative
to all LO QCD $2\to 2$ processes.}\label{fig:hadron}
\end{figure}

The only way to handle these complex processes is via Monte Carlo event
simulation. We have therefore implemented our cross section for elastic
parton-parton scattering via colour singlet exchange as a new hard subprocess in
the event generator {\sc pythia} \cite{Pythia}. Thus, higher order parton
emissions are included through conventional parton showers and the underlying
event is modelled with multiple interactions (using the best MI model {\tt
MSTP(82)=4}). To be consistent with our description \cite{SCI-TeV} of rapidity
gaps between a hard system ($W$, jets, bottom, $J/\psi$) and a leading proton
observed at the Tevatron, we also include the soft colour interaction (SCI)
model \cite{SCI}. The combination of SCI, with colour exchange probability
$P=0.25$, and MI used here to model the underlying event is consistent with our
previous study \cite{SCI-TeV}. Although SCI was introduced to explain the
creation of forward rapidity gaps, it has here the opposite effect by
rearranging strings to span across the gap produced at the parton level by the
BFKL singlet exchange. 

The gap fraction resulting from this complete model describes quite well 
the D\O\
data, as shown in fig.~\ref{fig:hadron}. The normalisation of the gap ratio is,
however, uncertain for two reasons. The parameter $s_0$ affects the
normalisation of the parton level cross section and the simulated survival
probability depends on the SCI model parameter $P$. A more complete study, 
including CDF data, will be presented in a coming paper \cite{coming}.

In conclusion, we have solved the nonforward BFKL equation for colour singlet
exchange including nonleading corrections. The part dominating at Tevatron
conditions is not the asymptotically dominating part contained in the
Mueller-Tang formula, but a part governed by a large logarithm that disappears
in the large-$y$ limit. Using {\sc pythia} with full event simulation we obtain
a dynamically varying gap survival probability resulting in a good description
of the Tevatron data on rapidity gaps between jets.

We thank J.~Kwieci\'{n}ski for stimulating discussions and acknowledge
the support from the Swedish Natural Science Research Council. LM is supported 
in part by the KBN grant number 5~P P03B 144 20.
\vspace{-2mm}


\end{document}